\documentclass[12pt]{article}
\usepackage{latexsym}
\usepackage{graphicx}
\usepackage{axodraw}
\usepackage{pst-plot,epsf}

\newcommand{\beq}{\begin{equation}}
\newcommand{\eeq}{\end{equation}}
\newcommand{\bea}{\begin{eqnarray}}
\newcommand{\eea}{\end{eqnarray}}
\newcommand{\nobody}{\rule{0ex}{1ex}}

\newcommand{\epm}{e^+e^-}
\newcommand{\gv}{\mbox{GeV}}

\newcommand{\ra}{\rightarrow}
\newcommand{\ga}{\gamma}
\newcommand{\eeudmn}{e^+ e^- \ra u \bar{d} \mu^- \bar{\nu}_{\mu}}

\begin{document}
\thispagestyle{empty}
\begin{flushright}
TP-USl/99/01\\
DESY 99-052\\
June 1999\\
\vspace*{1.5cm}
\end{flushright}
\begin{center}
{\LARGE\bf The hard bremsstrahlung correction \\ to
           $e^+ e^- \ra 4f$ with finite fermion masses: \\[2mm] results for
           $e^+ e^- \ra u \bar{d} \mu^- \bar{\nu}_{\mu}$\footnote{Work
supported in part by the Polish State Committee for Scientific
Research (KBN) under contract number 2~P03B~033~14.}
}\\
\vspace*{2cm}
Fred Jegerlehner$^{\rm a}$ and
Karol Ko\l odziej$^{\rm b}$\vspace{0.5cm}\\
$\nobody^{\rm a}${\small\it
Deutsches Elektronen-Synchrotron DESY, Platanenallee 6, D-15768 Zeuthen,
Germany}\\
$\nobody^{\rm b}${\small\it
Institute of Physics, University of Silesia, ul. Uniwersytecka 4,
PL-40007 Katowice, Poland}
\vspace*{3.5cm}\\
{\bf Abstract}\\
\end{center}
{\small An improved efficient method of calculating the hard
bremsstrahlung correction to $e^+ e^- \ra 4f$ for non-zero fermion
masses is presented. The non-vanishing fermion masses allow us to
perform the phase space integrations to the very collinear limit.  We
therefore can calculate cross sections independent of angular
cuts. Such calculations are important for background studies. Results
are presented for the total and some differential cross sections for
$e^+ e^- \ra u \bar{d} \mu^- \bar{\nu}_{\mu}$ and the corresponding hard
bremsstrahlung process. The latter is of particular interest for a
detailed investigation of the effects of final state radiation.
In principle, the process $e^+ e^- \ra u \bar{d} \mu^- \bar{\nu}_{\mu}\gamma$
is also interesting since it helps to set bounds on possible anomalous triple
and quartic gauge boson couplings involving photons. The size of
mass effects is illustrated by comparing the final states 
$u \bar{d} \mu^- \bar{\nu}_{\mu}(\gamma)$,
$c \bar{s} \mu^- \bar{\nu}_{\mu}(\gamma)$ and
$u \bar{d} \tau^- \bar{\nu}_{\tau}(\gamma)$.  }  

\vfill

\newpage

\section{Introduction}
\label{sec:1}

Precision measurements of properties of the intermediate gauge bosons
$Z$ and $W$ have deepened our understanding of the electroweak
interactions and consolidated the validity of the electroweak Standard
Model (SM) considerably in the past decade. After the convincing
success of LEP1 and SLD experiments in pinning down the properties of
the $Z$-resonance we expect further advances in measurements of the
properties of the $W$ boson which are not yet known with comparable
precision. As the SM very precisely predicts the mass and the width of
the $W$ a high accuracy determination of these parameters is of crucial
importance, because they allow us to improve indirect bounds on the
Higgs mass or on new physics beyond the SM. The precision of ongoing
measurements, single-$W$ production at the hadron collider TEVATRON,
and $W$-pair production at LEP2 are limited by statistics and/or by
lack of detailed theoretical understanding.

The proper analysis of $W^{\pm}$-pair production at LEP2 and later at
future high energy $e^+e^-$ linear colliders requires the accurate
knowledge of the SM predictions including all relevant radiative
corrections. What we need is a detailed understanding of the
production and subsequent decay of $W$ pairs, including the background
processes and photon radiation: $\epm \ra 4f,\,4f\gamma,4f
\gamma\gamma,\cdots$, where $4f$ denotes a possible four fermion final
state. The lowest order theoretical results for all the possible
four-fermion final states have been already implemented in several
Monte Carlo event generators and semi-analytic programs, which have
been thoroughly compared in \cite{Bardin}. Most of the programs
include some classes of radiative corrections such as the initial-
and final-state radiation, Coulomb corrections, running of the fine
structure constant, etc.  While presently available $\epm \ra
4f,\,4f\gamma$ matrix elements are precise enough for the analysis of
LEP2 data~\cite{LEP}, at future linear colliders, a much better
knowledge of the radiative corrections will be necessary because of
the high statistics expected at these accelerators and because
radiative corrections get more significant at higher energies.

The complete one-loop electroweak radiative corrections to the
on-shell $W^{\pm}$ pair production including soft bremsstrahlung were
calculated in \cite{rc}. The hard bremsstrahlung process $e^+e^-
\ra W^+W^-\gamma$ was included in \cite{KZ} and \cite{hb}. For
the process $e^+ e^- \ra W^+W^- \ra 4f$ of actual interest to the
experiments only partial results are available. We refer to
\cite{SD} for a recent review of the status of precision
calculations for this case.

Sufficiently above the $W^{\pm}$ pair production threshold,
for most of the present applications, it seems to be sufficient to
take into account corrections to the double-resonant diagrams only,
i.e., $e^+ e^- \ra 4f$ via virtual $W^+W^-$ intermediate states.
The validity of this approximation has to be controlled by more
complete calculations, however. From a theoretical point of view it is
certainly necessary to evaluate the complete $O(\alpha)$ radiative
corrections for the different channels of the $2\, \ra\, 4$ fermion
reactions. However, despite of the fact that some progress in
calculating the complete virtual one-loop electroweak radiative
corrections to $e^+ e^- \ra u \bar{d} \mu^- \bar{\nu}_{\mu}$ have been
reported in \cite{AV}, the final result of such a calculation is
still missing. Concerning the real photon emission, the situation
looks much better. The hard bremsstrahlung for four-fermion reactions
mediated by two resonant $W$ bosons was calculated in \cite{AW91}.
A similar calculation, extended by an inclusion of collinear
effects, was presented as a package {\tt WWF}~\cite{GJ}.
The complete lowest order result for $e^+ e^- \ra
e^-\bar{\nu}_e u \bar{d} \gamma$ was presented in \cite{JF} and
calculations of $e^+ e^- \ra 4f\gamma$ for an arbitrary final state
were reported in \cite{CM}. Results on bremsstrahlung for purely 
leptonic reactions have been published in \cite{SD} and 
most recently predictions for all processes $e^+ e^- \ra 4f\gamma$ with
massless fermions have been presented in \cite{DDRW}.

At a future linear collider, the proper treatment of the collinear
photons will be crucial and it requires to take into account the fermion
masses appropriately. Therefore, in the present paper, we propose an
efficient method of calculating the hard photon bremsstrahlung for
four fermion production in $e^+e^-$ annihilation without neglecting
the fermion masses. The phase space integration can therefore be
performed to the very collinear limit. This allows for calculating
cross sections independent of angular cuts and estimating background
contributions coming from undetected hard photons. We present results
for the total and a few differential cross sections for the channel
$e^+ e^-\ra u \bar{d} \mu^- \bar{\nu}_{\mu}$ and the corresponding
bremsstrahlung process. The latter is
particularly suited for a detailed investigation of effects related to
final state photon emission, since the muons appear well separated
from photons in the detectors. In particular it seems to be
interesting to study the influence of final state radiation on the $W$
mass measurement via this channel.
Having the final state photon resolution in
$e^+ e^-\ra u \bar{d} \mu^- \bar{\nu}_{\mu} \gamma$ could also make
it possible to investigate the quartic $\gamma VWW$ couplings
($V=\gamma,Z$), which are absent on the Born level of $4f$ production.
Of course, besides the new quartic
couplings there are additional triple $\gamma WW$ vertices as well.
In the soft photon limit, we can perform the integration over the soft
photon phase space analytically and demonstrate the cut-off independence
of the combined soft and hard photon bremsstrahlung cross section.
We finally will illustrate the importance of mass effects by comparing the
channels where $u\bar{d}$ is replaced by $c\bar{s}$. Similarly,
we may replace the $\mu$ by a $\tau$ lepton.

\section{Method of calculation}
\label{sec:2}

In this section, we present a method for calculating the matrix
elements of a two-fermion to four-fermion reaction and an associated
bremsstrahlung photon. The method is an extension of the helicity
amplitude method introduced in \cite{KZ} to final states of
arbitrary spin.

As in \cite{KZ}, we use the Weyl representation for fermions
where the Dirac matrices $\gamma^{\mu}, \mu=0,1,2,3,$ are given in
terms of the unit $2 \times 2$ matrix $I$ and Pauli matrices
$\sigma_i, i=1,2,3$, by \bea
\label{Weyl}
\gamma^{\mu}= \left( \begin{array}{cc}
                     0              & \sigma^{\mu}_+ \\
                 \sigma^{\mu}_- & 0
\end{array} \right),
\eea
with $\sigma^{\mu}_{\pm}=(I,\pm\sigma_i)$.
In representation (\ref{Weyl}), the matrix $\gamma_5 = i \gamma^0 \gamma^1
\gamma^2 \gamma^3$ and the chiral projectors $P_{\pm}=(1 \pm \gamma_5)/2$ read
\bea
\label{chiral}
\begin{array}{ccc}
\gamma_5= \left( \begin{array}{cc}
                -I & 0 \\
                 0 & I
\end{array} \right), \qquad
P_-= \left( \begin{array}{cc}
  I & 0 \\
  0 & 0
\end{array} \right), \qquad
P_+= \left( \begin{array}{cc}
  0 & 0 \\
  0 & I
\end{array} \right)
\end{array}.
\eea
A contraction of any four-vector $a^{\mu}$ with the $\gamma^{\mu}$ matrices
of (\ref{Weyl}) has the form
\bea
\label{contr}
/\!\!\!a = a^{\mu}\gamma_{\mu} = \left( \begin{array}{cc}
 0  & a^{\mu}\sigma_{\mu}^+ \\
a^{\mu}\sigma_{\mu}^- & 0
\end{array} \right)= \left( \begin{array}{cc}
 0  & a^+ \\
a^- & 0
\end{array} \right).
\eea
The $2 \times 2$ matrices $a^{\pm}$ can be expressed
in terms of the components of the four-vector $a^{\mu}$ by
\bea
\label{aba}
\begin{array}{cc}
a^+= \left( \begin{array}{cc}
 a^0 - a^3   & -a^1 + i a^2 \\
-a^1 - i a^2 &  a^0 + a^3
\end{array} \right), \qquad
a^-= \left( \begin{array}{cc}
 a^0 + a^3   & a^1 - i a^2 \\
 a^1 + i a^2 &  a^0 - a^3
\end{array} \right).
\end{array}
\eea

In representation (\ref{Weyl}), the helicity spinors for a particle,
$u({\bf p},\lambda)$, and an antiparticle, $v({\bf p},\lambda)$, of
four-momentum $(E, {\bf p})$ and helicity $\lambda/2=\pm 1/2$ are given by
\bea
\label{spinor}
\begin{array}{cc}
u({\bf p},\lambda) = \left( \begin{array}{c}
\sqrt{E-\lambda |{\bf p}|}\; \chi({\bf p},\lambda) \\
\sqrt{E+\lambda |{\bf p}|}\; \chi({\bf p},\lambda)
\end{array} \right), \qquad
v({\bf p},\lambda) = \left( \begin{array}{r}
-\lambda \sqrt{E+\lambda |{\bf p}|}\; \chi({\bf p},-\lambda) \\
\lambda \sqrt{E-\lambda |{\bf p}|}\; \chi({\bf p},-\lambda)
\end{array} \right)
\end{array},
\eea
and the helicity eigenstates $\chi({\bf p},\lambda)$ can be expressed
in terms of the spherical angles $\theta$ and $\phi$ of the momentum
vector ${\bf p}$ as\footnote{Note that our phase convention
differs from the one chosen in \cite{KZ}.}
\bea
\label{eigen}
\begin{array}{cc}
\chi({\bf p},+1) = \left( \begin{array}{r}
\cos{\theta/2} \\
e^{i\phi}\sin{\theta/2}
\end{array} \right), \qquad
\chi({\bf p},-1) = \left( \begin{array}{r}
-e^{-i\phi}\sin{\theta/2}\\
\cos{\theta/2}
\end{array} \right)
\end{array}.
\eea

For simplicity we use real polarization vectors which are defined again 
in terms of
$\theta$ and $\phi$
\bea
\label{polar}
\varepsilon^{\mu}({\bf p},1) = \left( 0,\cos{\theta}\cos{\phi},
\cos{\theta}\sin{\phi},-\sin{\theta} \right), \qquad
\varepsilon^{\mu}({\bf p},2) = \left( 0, -\sin{\phi}, \cos{\phi},0 \right) \\
\label{polar3}
\varepsilon^{\mu}({\bf p},3) = \gamma \left( \beta,\sin{\theta}\cos{\phi},
  \sin{\theta}\sin{\phi},\cos{\theta} \right),~~~~~~~~~~~~~~~~~~~~~~~~~~~~~~~
\eea
where the longitudinal polarization component of (\ref{polar3}) is
defined exclusively for a massive vector particle of energy $m\gamma$
and momentum $m \gamma \beta$. We could use complex polarization
vectors in the helicity basis as well, if we were interested in
definite helicity polarizations.

A polarized matrix element is calculated for a given set of external
particle momenta in a fixed reference frame, e.g.  the center of mass
system (c.m.s.) of the initial particles, where the initial momenta are
parallel to the $z$ axis.

Fermion masses are kept nonzero. The mass effects play an
essential role in the bremsstrahlung reactions whenever a collinear
photon is emitted. They also are important for tree level
reactions with identical particles in the initial and final state, where
a photon exchanged in the $t$-channel approaches its mass-shell.
Moreover, by keeping the fermion masses finite the Higgs boson exchange
can be incorporated in a consistent way.

In order to speed up numerical computation, we decompose the Feynman
graphs into factors which depend on a single uncontracted vector-index
and therefore may be considered as generalized polarization
vectors. They can be easily computed and used as building blocks of
other graphs.  For example, the coupling of an internal gauge boson to
the external fermions may be considered as the generalized
polarization vector $\varepsilon_V^{\mu}$ which is defined as
\bea
\label{polgen}
\varepsilon_V^{\mu}\left( p_1,p_2,\lambda_1,\lambda_2\right)=
D_V^{\mu\nu}(q) \bar{\psi}_1(p_1,\lambda_1)\gamma_{\nu}
\left( g_V^{(-)}P_- + g_V^{(+)}P_+ \right) \psi_2(p_2,\lambda_2)\nonumber \\
=\left[- \left(g_V^{(+)}\bar{\psi}_1^I\sigma^{\mu}_+\psi_2^{II} +
g_V^{(-)}\bar{\psi}_1^{II}\sigma^{\mu}_-\psi_2^{I}\right)
+ q^{\mu}/M_V^2\left( \left(m_1 g_V^{(-)} - m_2 g_V^{(+)}\right)
\bar{\psi}_1^I {\psi}_2^I  \right.\right.\nonumber \\  \left.\left.
+ \left(m_1 g_V^{(+)} - m_2 g_V^{(-)}\right)\bar{\psi}_1^{II}
{\psi}_2^{II}\right)\right]/\left(q^2 - M_V^2\right)
\eea
where $\bar{\psi}_1(p_1,\lambda_1)=(\bar{\psi}_1^I,\bar{\psi}_1^{II})$
and $\psi_2(p_2,\lambda_2)=({\psi}_2^I,{\psi}_2^{II})$ are spinors
for a particle or an antiparticle
of four-momentum $p_i$, mass $m_i$  and helicity $\lambda_i$,
as defined in (\ref{spinor}). We have denoted the chiral couplings of
the $\bar{\psi}_1\psi_2 V$ vertex by $ g_V^{(\pm)}$,
$D_V^{\mu\nu}(q)$ is the photon propagator in the Feynman gauge or
the massive gauge boson propagator in the unitary gauge
and $q=\pm p_1 \pm p_2$ is the four-momentum transfer. The $+(-)$-sign
corresponds to an outgoing (incoming) particle.
In the case of a photon propagator we have $M_V=0$ and only the first term
in the square brackets on the right hand side of (\ref{polgen})
is present.


The photon emission from any of the external fermion legs
of the $\bar{\psi}_1\psi_2 V$ vertex can be taken into account by
defining two other generalized polarization vectors:
\bea
\label{polgen1}
\varepsilon_{\gamma V}^{\mu}\left(p_1,p_2,k,\lambda_1,\lambda_2,\lambda
\right) = D_V^{\mu\nu}(q+k)\;\bar{\psi}_1(p_1,\lambda_1)
g_{\gamma 1} /\!\!\!{\varepsilon}(k,\lambda)
{{\pm /\!\!\!p_1 + /\!\!\!k + m_1}\over{(\pm p_1 + k)^2 - m_1^2}}\nonumber \\
\times \quad \gamma_{\nu} \left( g_V^{(-)}P_- + g_V^{(+)}P_+ \right)
\psi_2(p_2,\lambda_2)\nonumber \\
={g_{\gamma 1} \over {2p_1\cdot k}}D_V^{\mu\nu}(q+k)
         \left[g_V^{(+)}\bar{\psi}_1^I \left( {2p_1\cdot \varepsilon}
                   \mp k^+\varepsilon^-\right)\sigma_{\nu}^+\psi_2^{II}
     +g_V^{(-)}\bar{\psi}_1^{II} \left( {2p_1\cdot \varepsilon}
        \mp k^-\varepsilon^+ \right)\sigma_{\nu}^-\psi_2^{I} \right]
\eea
where the upper (lower) sign is assumed if $\psi_1$ represents an outgoing
particle (incoming antiparticle) and
\bea
\label{polgen2}
\varepsilon_{V\gamma}^{\mu}\left(p_1,p_2,k,\lambda_1,\lambda_2,\lambda
\right) = D_V^{\mu\nu}(q+k)\;\bar{\psi}_1(p_1,\lambda_1)
\gamma_{\nu} \left( g_V^{(-)}P_- + g_V^{(+)}P_+ \right) \nonumber \\
\times \quad {{\pm/\!\!\!p_2 - /\!\!\!k + m_2}\over{(\pm p_2 - k)^2 - m_2^2}}
g_{\gamma 2} /\!\!\!{\varepsilon}(k,\lambda)\psi_2(p_2,\lambda_2)\nonumber \\
={g_{\gamma 2} \over {2p_2\cdot k}}D_V^{\mu\nu}(q+k)
         \left[ g_V^{(+)}\bar{\psi}_1^I \sigma_{\nu}^+
\left(-{2p_2\cdot \varepsilon}
                        \pm k^-\varepsilon^+\right)\psi_2^{II}
     +g_V^{(-)}\bar{\psi}_1^{II} \sigma_{\nu}^-\left(-{2p_2\cdot \varepsilon}
        \pm k^+\varepsilon^- \right)\psi_2^{I} \right]
\eea
where the upper (lower) sign has to be taken when $\psi_2$ represents an
incoming particle (outgoing antiparticle). In
(\ref{polgen1}) and (\ref{polgen2}),
$\varepsilon^{\mu}(k,\lambda)$ is the photon polarization vector
as defined in (\ref{polar}) and $g_{\gamma i}$ are the
photon couplings to $\psi_i$.

If we contract the triple gauge boson coupling
\bea
\label{triple}
\Gamma_{(WWV)}^{\mu\nu\rho}(p_1,p_2,p_3)=
g_{WWV}\left[ (p_1-p_2)^{\rho}g^{\mu\nu}+(p_2-p_3)^{\mu}g^{\nu\rho}
+(p_3-p_1)^{\nu}g^{\mu\rho}\right],
\eea
where $p_1, p_2$ and $p_3$ are the incoming momenta of the $W_{\mu}^+,
W_{\nu}^-$ and the neutral gauge boson $V_{\rho}$, $V = \gamma, Z^0$,
respectively, with two (generalized) polarization vectors,
say $\varepsilon^{\nu}_1$, $\varepsilon^{\rho}_2$, and
with a gauge boson propagator
we will obtain another
generalized polarization vector, e.g.
\bea
\label{polgen3}
\varepsilon_{V}^{\sigma}(1,2)=D_V^{\sigma\mu}(q)
\Gamma_{\mu\nu\rho}^{(WWV)} \varepsilon_1^{\nu}\varepsilon_2^{\rho}.
\eea
In (\ref{polgen3}), 1 and 2 stay for the four-momenta and polarizations.
With the help of the generalized polarization vectors
(\ref{polgen}--\ref{polgen2}) and (\ref{polgen3}), the amplitude
corresponding to any Feynman diagram of a process $e^+e^- \ra 4f(\gamma)$
may be represented by one of the scalar functions $F_{2n+1}$, $E_3$
or $E_4$ we are going to define now.
A fermion line containing $n+1$ couplings to gauge bosons and $n$
fermion propagators sandwiched between external spinors can be represented
by the scalar function
\bea
\label{f}
F_{2n+1}\left(\bar{\psi}_1,g_1^{(+)}\varepsilon_1^+,g_1^{(-)}\varepsilon_1^-,
p_1^+,p_1^-,m_1,g_2^{(+)}\varepsilon_2^+,g_2^{(-)}\varepsilon_2^-,
p_2^+,p_2^-,m_2,\cdots, \right. \nonumber \\
\left. p_n^+,p_n^-,m_n,g_{n+1}^{(+)}\varepsilon_{n+1}^+,
g_{n+1}^{(-)}\varepsilon_{n+1}^-,\psi_2\right)= \nonumber \\
\bar{\psi}_1 /\!\!\!{\varepsilon_1} \left( g_1^{(-)}P_- + g_1^{(+)}P_+ \right)
{{/\!\!\!p_1 + m_1}\over{p_1^2 - m_1^2}}
/\!\!\!{\varepsilon_2} \left( g_2^{(-)}P_- + g_2^{(+)}P_+ \right)
{{/\!\!\!p_2 + m_2}\over{p_2^2 - m_2^2}}\cdots \nonumber \\
\times {{/\!\!\!p_n + m_n}\over{p_n^2 - m_n^2}}  /\!\!\!{\varepsilon_{n+1}}
\left( g_{n+1}^{(-)}P_- + g_{n+1}^{(+)}P_+ \right) \psi_2,
\eea
where we have suppressed polarization indices. In the representation
(\ref{Weyl}--\ref{spinor}), the algebra of $4 \times 4$ matrices in
(\ref{f}) can be easily reduced to the algebra of $2 \times 2$
matrices. This kind of reduction has been used already in
(\ref{polgen}--\ref{polgen2}). Utilizing the $2 \times 2$
algebraic representation speeds up the numerical computation and
allows for a simultaneous calculation of the $\gamma$ and $Z$
contributions. We have indicated the use of the reduced form by
writing explicitly the dependence on the $2 \times 2$ matrices
$\varepsilon_i^{\pm}, p_i^{\pm}$, defined according to
(\ref{aba}), on the left hand side of (\ref{f}). The general
form of this reduction is quite a complicated formula. Therefore we
restrict ourselves to present an example of the function $F_3$ which can be
written as
\bea
\label{f3}
F_3\left(\bar{\psi}_1,g_1^{(+)}\varepsilon_1^+,g_1^{(-)}\varepsilon_1^-,
p_1^+,p_1^-,m_1,g_2^{(+)}\varepsilon_2^+,g_2^{(-)}\varepsilon_2^-\right)
= m_1 g_1^{(+)}g_2^{(-)}
        \bar{\psi}_1^{I}\varepsilon_1^+\varepsilon_2^-\psi_2^{I} \nonumber \\
+g_1^{(+)}g_2^{(+)}
        \bar{\psi}_1^{I}\varepsilon_1^+p_1^-\varepsilon_2^+\psi_2^{II}
+g_1^{(-)}g_2^{(-)}
        \bar{\psi}_1^{II}\varepsilon_1^- p_1^+ \varepsilon_2^-\psi_2^{I}
+m_1 g_1^{(-)}g_2^{(+)}
        \bar{\psi}_1^{II}\varepsilon_1^-\varepsilon_2^+\psi_2^{II}.
\eea
A contraction of the triple gauge boson coupling (\ref{triple})
with three polarization vectors
$\varepsilon_1^{\mu}, \varepsilon_2^{\nu} $ and $ \varepsilon_3^{\rho}$
 can be considered as a scalar function
\bea
\label{e3}
E_{3}\left[p_1,\varepsilon_1,p_2,\varepsilon_2,p_3,
                     \varepsilon_3 \right)&=&          \nonumber \\
g_{WWV}& & \hspace*{-1.cm}\left(
 (p_1-p_2)\!\cdot \!\varepsilon_3 \; \varepsilon_1 \!\cdot\! \varepsilon_2
+(p_2-p_3)\!\cdot\! \varepsilon_1 \;\varepsilon_2 \!\cdot\! \varepsilon_3
+(p_3-p_1)\!\cdot\! \varepsilon_2 \;\varepsilon_1 \!\cdot\! \varepsilon_3
                                                                    \right].
\eea
Similarly, the quartic gauge boson coupling
\bea
\label{quartic}
\Gamma_{(4)}^{\mu\nu\rho\sigma} = g_{V_1V_2WW}
\left( g^{\mu\rho}g^{\nu\sigma}+g^{\mu\sigma}g^{\nu\rho}
      -2 g^{\mu\nu}g^{\rho\sigma}\right),
\eea
where the vector indices $\mu, \nu$ are associated with the neutral
gauge bosons $V_1$ and $V_2$ and $\rho, \sigma$ with $W^+$ and $W^-$,
if contracted with four polarization vectors $\varepsilon_1^{\mu},
\varepsilon_2^{\nu}, \varepsilon_3^{\rho}$ and
$\varepsilon_4^{\sigma}$ can be considered as another scalar function
\bea
\label{e4}
E_{4}\left(\varepsilon_1,\varepsilon_2,\varepsilon_3,\varepsilon_4 \right)
=g_{V_1V_2WW} \left(
\varepsilon_1 \!\cdot \!\varepsilon_3 \; \varepsilon_2 \!\cdot\! \varepsilon_4
+\varepsilon_1 \!\cdot \!\varepsilon_4 \; \varepsilon_2 \!\cdot\! \varepsilon_3
-2
\varepsilon_1 \!\cdot \!\varepsilon_2 \; \varepsilon_3 \!\cdot\! \varepsilon_4
                                                                       \right).
\eea
Note that similar scalar functions can be defined for the Higgs boson by
replacing the vector boson coupling  and propagator in
(\ref{polgen}--\ref{polgen2}) and (\ref{polgen3}) by the Higgs
coupling and propagator.

Functions (\ref{f}), (\ref{e3}) and (\ref{e4}) can easily be implemented in
a {\tt FORTRAN} program and computed numerically for any specific set
of the particle momenta and polarizations. The Fortran 90
language standard which contains a number of new features and
intrinsic functions especially for array manipulations is particularly
suitable for this task.

The method described above can be used to calculate the matrix
element of any process of
$e^+e^-$ annihilation into four fermions and a photon.
Actually, the method is quite general and can be applied to any $2 \ra n$
tree level reaction, not necessarily in the framework of the standard model,
with massive or massless fermions and/or bosons in the final state.
However, it may happen that one or a few extra functions  will have
to be defined in addition to those defined in (\ref{f}), (\ref{e3})
and (\ref{e4}).
Practically the only limitation of the method is the feasibility of
phase space integration which is performed numerically by applying the Monte
Carlo method.

\section{Application to $e^+ e^- \ra u \bar{d} \mu^- \bar{\nu}_{\mu}$
and $e^+ e^- \ra u \bar{d} \mu^- \bar{\nu}_{\mu}\gamma$}
\label{sec:3}

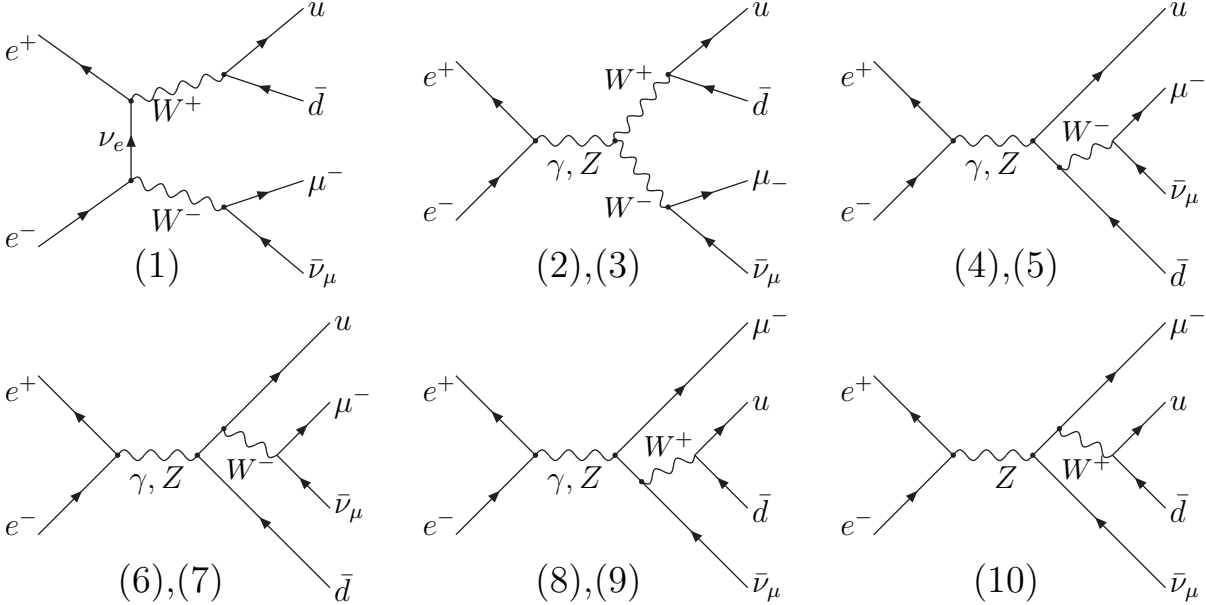
\begin{figure}[htb]
\centerline{
\begin{picture}(120,100)(10,0)
\ArrowLine(10,10)(45,35)
\put(10,10){\makebox(0,0)[br]{$e^-$}}
\ArrowLine(45,35)(45,65)
\put(43,50){\makebox(0,0)[r]{$\nu_e$}}
\ArrowLine(45,65)(10,90)
\put(10,90){\makebox(0,0)[tr]{$e^+$}}
\Vertex(45,35){1}
\Vertex(45,65){1}
\Photon(45,65)(80,75){2}{4}
\put(62.5,67){\makebox(0,0)[t]{$W^+$}}
\Vertex(80,75){1}
\ArrowLine(110,65)(80,75)
\put(112,65){\makebox(0,0)[l]{$\bar{d}$}}
\ArrowLine(80,75)(110,100)
\put(112,100){\makebox(0,0)[l]{$u$}}
\Photon(45,35)(80,25){2}{4}
\put(62.5,25){\makebox(0,0)[t]{$W^-$}}
\Vertex(80,25){1}
\ArrowLine(110,0)(80,25)
\put(112,0){\makebox(0,0)[l]{$\bar{\nu}_{\mu}$}}
\ArrowLine(80,25)(110,35)
\put(112,35){\makebox(0,0)[l]{$\mu^-$}}
\put(55,-5){\makebox(0,0)[b]{\large (1)}}
\end{picture}
\hfill
\begin{picture}(120,100)(0,0)
\ArrowLine(0,20)(30,50)
\put(0,20){\makebox(0,0)[br]{$e^-$}}
\ArrowLine(30,50)(0,80)
\put(0,80){\makebox(0,0)[tr]{$e^+$}}
\Vertex(30,50){1}
\Photon(30,50)(60,50){2}{3}
\put(45,46){\makebox(0,0)[t]{$\gamma,Z$}}
\Vertex(60,50){1}
\Photon(60,50)(80,75){2}{4}
\put(75,70){\makebox(0,0)[br]{\small$W^+$}}
\Vertex(80,75){1}
\ArrowLine(110,65)(80,75)
\put(112,65){\makebox(0,0)[l]{$\bar{d}$}}
\ArrowLine(80,75)(110,100)
\put(112,100){\makebox(0,0)[l]{$u$}}
\Photon(60,50)(80,25){2}{4}
\put(75,30){\makebox(0,0)[tr]{\small$W^-$}}
\Vertex(80,25){1}
\ArrowLine(110,0)(80,25)
\put(112,0){\makebox(0,0)[l]{$\bar{\nu}_{\mu}$}}
\ArrowLine(80,25)(110,35)
\put(112,35){\makebox(0,0)[l]{$\mu_-$}}
\put(50,-5){\makebox(0,0)[b]{\large (2),(3)}}
\end{picture}
\hfill
\begin{picture}(120,100)(0,0)
\ArrowLine(0,20)(30,50)
\put(0,20){\makebox(0,0)[br]{$e^-$}}
\ArrowLine(30,50)(0,80)
\put(0,80){\makebox(0,0)[tr]{$e^+$}}
\Vertex(30,50){1}
\Photon(30,50)(60,50){2}{3}
\put(45,46){\makebox(0,0)[t]{$\gamma,Z$}}
\Vertex(60,50){1}
\ArrowLine(110,0)(70,40)
\put(112,0){\makebox(0,0)[l]{$\bar{d}$}}
\Line(70,40)(60,50)
\Vertex(70,40){1}
\Photon(70,40)(90,50){-2}{2.5}
\put(90,50){\makebox(0,0)[br]{\small$W^-$}}
\ArrowLine(110,30)(90,50)
\put(112,30){\makebox(0,0)[l]{$\bar{\nu}_{\mu}$}}
\ArrowLine(90,50)(110,70)
\put(112,70){\makebox(0,0)[l]{$\mu^-$}}
\ArrowLine(60,50)(110,100)
\put(112,100){\makebox(0,0)[l]{$u$}}
\put(50,-5){\makebox(0,0)[b]{\large (4),(5)}}
\end{picture}
\hfill
}
\bigskip
\medskip
\centerline{
\begin{picture}(120,100)(0,0)
\ArrowLine(0,20)(30,50)
\put(0,20){\makebox(0,0)[br]{$e^-$}}
\ArrowLine(30,50)(0,80)
\put(0,80){\makebox(0,0)[tr]{$e^+$}}
\Vertex(30,50){1}
\Photon(30,50)(60,50){2}{3}
\put(45,46){\makebox(0,0)[t]{$\gamma,Z$}}
\Vertex(60,50){1}
\ArrowLine(110,0)(60,50)
\put(112,0){\makebox(0,0)[l]{$\bar{d}$}}
\Line(60,50)(70,60)
\Vertex(70,60){1}
\Photon(70,60)(90,50){-2}{2.5}
\put(90,50){\makebox(0,0)[tr]{\small$W^-$}}
\ArrowLine(110,30)(90,50)
\put(112,30){\makebox(0,0)[l]{$\bar{\nu}_{\mu}$}}
\ArrowLine(90,50)(110,70)
\put(112,70){\makebox(0,0)[l]{$\mu^-$}}
\ArrowLine(70,60)(110,100)
\put(112,100){\makebox(0,0)[l]{$u$}}
\put(50,-5){\makebox(0,0)[b]{\large (6),(7)}}
\end{picture}
\hfill
\begin{picture}(120,100)(0,0)
\ArrowLine(0,20)(30,50)
\put(0,20){\makebox(0,0)[br]{$e^-$}}
\ArrowLine(30,50)(0,80)
\put(0,80){\makebox(0,0)[tr]{$e^+$}}
\Vertex(30,50){1}
\Photon(30,50)(60,50){2}{3}
\put(45,46){\makebox(0,0)[t]{$\gamma,Z$}}
\Vertex(60,50){1}
\ArrowLine(110,0)(70,40)
\put(112,0){\makebox(0,0)[l]{$\bar{\nu}_{\mu}$}}
\Line(70,40)(60,50)
\Vertex(70,40){1}
\Photon(70,40)(90,50){-2}{2.5}
\put(90,50){\makebox(0,0)[br]{\small$W^+$}}
\ArrowLine(110,30)(90,50)
\put(112,30){\makebox(0,0)[l]{$\bar{d}$}}
\ArrowLine(90,50)(110,70)
\put(112,70){\makebox(0,0)[l]{$u$}}
\ArrowLine(60,50)(110,100)
\put(112,100){\makebox(0,0)[l]{$\mu^-$}}
\put(50,-5){\makebox(0,0)[b]{\large (8),(9)}}
\end{picture}
\hfill
\begin{picture}(120,100)(0,0)
\ArrowLine(0,20)(30,50)
\put(0,20){\makebox(0,0)[br]{$e^-$}}
\ArrowLine(30,50)(0,80)
\put(0,80){\makebox(0,0)[tr]{$e^+$}}
\Vertex(30,50){1}
\Photon(30,50)(60,50){2}{3}
\put(50,46){\makebox(0,0)[t]{$Z$}}
\Vertex(60,50){1}
\ArrowLine(110,0)(60,50)
\put(112,0){\makebox(0,0)[l]{$\bar{\nu}_{\mu}$}}
\Line(60,50)(70,60)
\Vertex(70,60){1}
\Photon(70,60)(90,50){-2}{2.5}
\put(90,50){\makebox(0,0)[tr]{\small$W^+$}}
\ArrowLine(110,30)(90,50)
\put(112,30){\makebox(0,0)[l]{$\bar{d}$}}
\ArrowLine(90,50)(110,70)
\put(112,70){\makebox(0,0)[l]{$u$}}
\ArrowLine(70,60)(110,100)
\put(112,100){\makebox(0,0)[l]{$\mu^-$}}
\put(50,-5){\makebox(0,0)[b]{\large (10)}}
\end{picture}
\hfill
}
\medskip
\caption[]{The Feynman diagrams of reaction (\ref{Born}).}
\label{fig:1}
\end{figure}

Let us demonstrate how the method of Sect.~\ref{sec:2} works in case of the tree level
four-fermion reaction
\bea
\label{Born}
  e^+(p_1,\lambda_1) + e^-(p_2,\lambda_2) \ra u(p_3,\lambda_3)
+ \bar{d}(p_4,\lambda_4) + \mu^-(p_5,\lambda_5)
                             + \bar{\nu}_{\mu}(p_6,\lambda_6),
\eea
where the four-momenta and helicities are indicated in parenthesis.
The Feynman diagrams of the process (\ref{Born}) are shown in
Fig.~\ref{fig:1}.  Although the fermion mass effects are irrelevant
for the reaction (\ref{Born}), we keep masses finite for the sake of
illustration.  However, we neglect the Higgs boson exchange
contribution which is suppressed by ratios of the fermion masses to
the $W$ boson mass.

We define the necessary generalized polarization vectors
using (\ref{polgen})
\bea
\label{pol}
\varepsilon_{\gamma}^{\mu}(p_1,p_2,\lambda_1,\lambda_2)&=&
D_{\gamma}^{\mu\nu}(p_{12})\;
\bar{v}_1(p_1,\lambda_1)\gamma_{\nu} g_{\gamma e} u_2(p_2,\lambda_2),
                                                              \nonumber \\
\varepsilon_Z^{\mu}(p_1,p_2,\lambda_1,\lambda_2)&=&
D_{Z}^{\mu\nu}(p_{12})\;
\bar{v}_1(p_1,\lambda_1) \gamma_{\nu}
     \left( g_{Ze}^{(-)}P_- + g_{Ze}^{(+)}P_+ \right)
                                           u_2(p_2,\lambda_2), \nonumber \\
\varepsilon_{W^+}^{\mu}(p_3,p_4,\lambda_3,\lambda_4)&=&
D_{W}^{\mu\nu}(p_{34})\;
\bar{u}_3(p_3,\lambda_3)\gamma_{\nu} g_W P_- v_4(p_4,\lambda_4), \nonumber \\
\varepsilon_{W^-}^{\mu}(p_5,p_6,\lambda_5,\lambda_6)&=&
D_{W}^{\mu\nu}(p_{56})\;
\bar{u}_5(p_5,\lambda_5)\gamma_{\nu} g_W P_- v_6(p_6,\lambda_6),
\eea
where we have introduced the shorthand notation $p_{12}=p_1+p_2$, 
$p_{34}=p_3+p_4$ and $p_{56}=p_5+p_6$; $g_{\gamma e}, g_{Z e}^{(\pm)}$ 
and $g_W$ are the standard model couplings.
We use constant widths for the massive
gauge bosons. They are introduced through the complex mass parameters
$M_V^2=m_V^2-im_V\Gamma_V$ in the propagators $D_V^{\mu\nu}$,$V=W,Z$. However,
we keep a real value of the electroweak mixing parameter $\sin\theta_W$.
This simple prescription preserves the electromagnetic gauge invariance,
also for the nonzero fermion masses. This has been checked
analytically and confirmed by the numerical calculation. We would like
to stress at this point that this result is obtained with two
independent widths $\Gamma_W$ and $\Gamma_Z$ which violate
the $SU(2)$ gauge invariance. This finding seems to contradict
the discussion of this issue in \cite{BD}.
The resulting violation of the high energy unitarity cancellations
for $e^+ e^- \ra 4f$ is suppressed by the factor $\Gamma_W M_W/s$. A comparison
of different gauge-boson width prescriptions performed in \cite{SD}
and \cite{DDRW} shows that our simple prescription is satisfactory as far 
as the experimental precision of LEP2 and future linacs is concerned.

Using polarization vectors (\ref{pol}) we can express the helicity
amplitudes of reaction (\ref{Born}) corresponding to diagrams of
Fig.~\ref{fig:1} in terms of the functions defined in (\ref{f}),
(\ref{e3}) and (\ref{e4})
\bea
\label{ampli}
M_1&=&F_3\left( \bar{v}_1,0,g_W\varepsilon_{W^+}^-,
      p_2^+-p_{56}^+,p_2^--p_{56}^-,0,g_W\varepsilon_{W^-}^-,u_2 \right),
                                                       \nonumber \\
M_2+M_3&=&E_3\left( p_{56},\varepsilon_{W^-},p_{34},\varepsilon_{W^+},
      p_{12},g_{WW\gamma}\varepsilon_{\gamma}+g_{WWZ}\varepsilon_{Z}\right)
                                                       \nonumber \\
M_4+M_5&=&F_3\left( \bar{u}_3,
   g_{\gamma u}{\varepsilon}_{\gamma}^+ +g_{Zu}^{(+)}{\varepsilon}_{Z}^+,
   g_{\gamma u}\varepsilon_{\gamma}^- +g_{Zu}^{(-)}\varepsilon_{Z}^-,
   p_3^+ -p_{12}^+,p_3^- -p_{12}^-,m_3,0,g_W\varepsilon_{W^-}^-,v_4
                                                    \right),\nonumber \\
M_6+M_7&=&F_3\left( \bar{u}_3,0,g_W\varepsilon_{W^-}^-,
      p_{12}^+-p_4^+,p_{12}^- -p_4^-,m_4,
g_{\gamma d} \varepsilon_{\gamma}^+ +g_{Zd}^{(+)} \varepsilon_{Z}^+,
g_{\gamma d} \varepsilon_{\gamma}^- +g_{Zd}^{(-)} \varepsilon_{Z}^-,
                                          v_4 \right),  \nonumber \\
M_8+M_9&=&F_3\left( \bar{u}_5,
 g_{\gamma\mu}{\varepsilon}_{\gamma}^+ +g_{Z\mu}^{(+)}{\varepsilon}_{Z}^+,
 g_{\gamma\mu}{\varepsilon}_{\gamma}^- +g_{Z\mu}^{(-)}{\varepsilon}_{Z}^-,
 p_5^+-p_{12}^+,p_5^- -p_{12}^-,m_5,0,g_W\varepsilon_{W^+}^-,v_6
                                            \right), \nonumber \\
M_{10}&=&F_3\left( \bar{u}_5,0,g_W\varepsilon_{W^+}^-,
      p_{12}^+ -p_6^+,p_{12}^- -p_6^-,0,
      0,g_{Z\nu}\varepsilon_{Z}^-,v_4 \right),
\eea
where we have used the shorthand
notation $v_i=v_i({\bf p}_i,\lambda_i),$ $i=1,4,6,$
$u_j=u_j({\bf p}_j,\lambda_j),$ $j=2,3,5$ for
the spinors which are defined according to (\ref{spinor}).
The standard model couplings of (\ref{pol}) and (\ref{ampli}) are
defined in terms of the electric charge $e$ and the electroweak mixing
parameter $\sin^2\theta_W$. 

By comparing (\ref{ampli}) with Fig.~\ref{fig:1} one observes that the
diagrams which differ only by the replacement of the photon and $Z$
propagators can be calculated simultaneously. This is one of the
advantages of the presented method.

Now the modulus squared of the spin averaged matrix element of the
reaction ({\ref{Born}) can be easily computed numerically.

The matrix element of the bremsstrahlung reaction
\bea
\label{brems}
  e^+(p_1) + e^-(p_2) \ra u(p_3) + \bar{d}(p_4) + \mu^-(p_5)
                          + \bar{\nu}_{\mu}(p_6) + \gamma(p_7),
\eea
where the particle four-momenta are indicated in parenthesis is
calculated in the same way.  The 71 Feynman diagrams of the process
(\ref{brems}) can be obtained from those of Fig.~\ref{fig:1} by
attaching an external photon line to each charged particle as well as
to the triple gauge boson vertex. We again neglect the Higgs boson
contribution.

In the soft photon limit, $|{\bf p}_7| < \omega$, the matrix element of 
reaction (\ref{brems}) takes the simple factorized form
\bea
\label{soft}
& &\left.M_{\gamma}(p_1,p_2,p_3,p_4,p_5,p_6,p_7,\lambda_1,\lambda_2,\lambda_3,
   \lambda_4,\lambda_5,\lambda_6,\lambda_7)\right|_{|{\bf p}_7| < \omega}=
                                                       \nonumber \\
&-&\left(  g_{\gamma l} {{p_1^{\mu}} \over {p_1 \cdot p_7}}
       - g_{\gamma l} {{p_2^{\mu}} \over {p_2 \cdot p_7}}
       + g_{\gamma u} {{p_3^{\mu}} \over {p_3 \cdot p_7}}
       - g_{\gamma d} {{p_4^{\mu}} \over {p_4 \cdot p_7}}
       + g_{\gamma l} {{p_5^{\mu}} \over {p_5 \cdot p_7}}\right) 
        \varepsilon_{\mu}(p_7,\lambda_7)\\
&\times & M_0(p_1,p_2,p_3,p_4,p_5,p_6,\lambda_1,\lambda_2,\lambda_3,
   \lambda_4,\lambda_5,\lambda_6), \nonumber 
\eea
where $M_0$ is the matrix element of reaction (\ref{Born}) and the 
photon-fermion couplings are given by
$g_{\gamma l}=e$, $g_{\gamma u}=2/3e$ and $g_{\gamma d}=-e/3$.

\section{The phase space integration}
\label{sec:4}

The phase space integration is performed with the Monte Carlo integration
routine {\tt VEGAS}~\cite{vegas}. We integrate out the dependence on
the azimuthal angle related to the rotational symmetry with respect to 
the beam axis. This symmetry is satisfied as long
as we do not consider transversely polarized initial beams.
Thus, the number of integrations is reduced from 8 to 7 and from
11 to 10 for reactions (\ref{Born}) and (\ref{brems}), respectively.

The 7 dimensional phase space element of the reaction (\ref{Born}) is
parametrized by
\bea
\label{ps7}
{\rm d}^7Lips&=&(2\pi)^{-7} \; \frac {\lambda^{1/2}(s,s_{34},s_{56})}{8s}\;
           \frac {\lambda^{1/2}(s_{34},m_3^2,m_4^2)}{8s_{34}}\;
           \frac {\lambda^{1/2}(s_{56},m_5^2,m_6^2)}{8s_{56}}
           \nonumber \\ &&~~\times ~~
{\rm d}s_{34} {\rm d}s_{56} {\rm d}\cos\theta {\rm d}\Omega_3 {\rm d}\Omega_5,
\eea
where $s=(p_1 + p_2)^2$, $s_{34}=(p_3 + p_4)^2$, $s_{56}=(p_5 +
p_6)^2$, $\theta$ is an angle between the momenta ${\bf p}_3 + {\bf
p}_4$ and the $z$ axis of the c.m. system which is
directed along the positron momentum ${\bf p}_1$.  ${\rm d}\Omega_3 =
{\rm d}\cos\theta_3{\rm d}\phi_3$ (${\rm d}\Omega_5 = {\rm
d}\cos\theta_5{\rm d}\phi_5$ ) is the solid angle element of ${\bf
p}_3$ (${\bf p}_5$) in the respective c.m. frame where ${\bf p}_3 +
{\bf p}_4 = 0$ (${\bf p}_5 + {\bf p}_6 = 0$).

The integration limits in the invariants $s_{34}$ and $s_{56}$
of (\ref{ps7}) are given by
\bea
\label{s034}
(m_3+m_4)^2 &\le & s_{34} \le (\sqrt{s}-m_5-m_6)^2, \\
\label{s056}
(m_5+m_6)^2 &\le & s_{56} \le (\sqrt{s}-\sqrt{s_{34}})^2
\eea
and the spherical angles vary in the full range, i.e.
\bea
\label{spheric}
 -1 & \le & \cos{\theta} \le 1,\nonumber \\
0 & \le & \Omega_i \le 4\pi, \quad i=3,5.
\eea

The 10 dimensional phase space element of the reaction (\ref{brems}) is
parametrized in different ways dependent on whether we want to account
for the peaking related to the photon emission from the initial or
final state fermions. In order to deal with the initial state radiation 
peaking we parametrize the phase space by
\bea
\label{ps10i}
{\rm d}^{10}Lips&=&(2\pi)^{-10} \; \frac{E_7}{2} \;
\frac {\lambda^{1/2}(s',s_{34},s_{56})}{8s'} \;
           \frac {\lambda^{1/2}(s_{34},m_3^2,m_4^2)}{8s_{34}}\;
           \frac {\lambda^{1/2}(s_{56},m_5^2,m_6^2)}{8s_{56}}
           \nonumber \\ &&~~\times ~~
{\rm d}E_7 {\rm d}\Omega_7 {\rm d}s_{34} {\rm d}s_{56}
{\rm d}\cos\theta_{34} {\rm d}\Omega_3 {\rm d}\Omega_5,
\eea
where $s'=(p_1 + p_2 - p_7)^2$. The photon variables, the energy $E_7$,
and the solid angle $\Omega_7$, are defined in the frame where
${\bf p}_1 + {\bf p}_2 - {\bf p}_7 = 0$. The polar angle $\theta_{34}$
of the momentum vector ${\bf p}_3 + {\bf p}_4$ with respect to the
positron beam is defined in the same frame. The invariant masses
$s_{34}$, $s_{56}$ and solid angles $\Omega_{3}$, $\Omega_{5}$
are defined as in (\ref{ps7}). 

The integration limits in the photon energy $E_7$ and in the invariants 
$s_{34}, s_{56}$ of (\ref{ps10i}) read
\bea
\label{e7}
E_{\rm cut} \le E_7 &\le& \left(s-(m_3+m_4+m_5+m_6)^2\right)/(2\sqrt{s}),\\
\label{s34}
(m_3+m_4)^2 \le s_{34} &\le& (\sqrt{s'}-m_5-m_6)^2, \\
\label{s56}
(m_5+m_6)^2 \le s_{56} &\le& (\sqrt{s'}-\sqrt{s_{34}})^2,
\eea
where $E_{\rm cut}$ is the minimum hard photon energy to be detected
and $s'=\sqrt{s}(\sqrt{s}-2E_7)$.
The spherical angles of (\ref{ps10i}) vary in the
full range, as in (\ref{spheric}).

On the other hand, the radiation off the final state $\bar{d}$ quark is 
dealt with another parametrization
\bea
\label{ps10f}
{\rm d}^{10}Lips &=& (2\pi)^{-10} 
\frac {\lambda^{1/2}(s,s_{347},s_{56})}{8s} \;
\frac {\lambda^{1/2}(s_{56},m_5^2,m_6^2)}{8s_{56}}\;\nonumber \\
&\times& \frac{1}{8} {\rm d}s_{347} {\rm d}s_{56}{\rm d}\cos\theta_{347}
{\rm d}E_3 {\rm d}E_7 {\rm d}\Omega_3 {\rm d}\phi_{37} 
{\rm d}\Omega_5,
\eea
where the polar angle $\theta_{347}$ of the momentum ${\bf p}_3+{\bf
p}_4+{\bf p}_7$ is defined in the c.m.s.; the energy $E_3$ and the
spherical angle $\Omega_3$ of the $u$ quark, the photon energy $E_7$
and the azimuthal angle $\phi_{37}$ between the $u$ quark and the
$\gamma$ are defined in the relative c.m.s. of $u$ and $\bar{d}$
quarks and the photon; the spherical angle $\Omega_5$ is defined in
the c.m.s. of the $\mu^-$ and $\bar{\nu}_{\mu}$.

The integration limits are now specified by
\bea
\label{s347}
(m_3+m_4)^2 \le s_{347} &\le& (\sqrt{s}-m_5-m_6)^2, \\
\label{s56f}
(m_5+m_6)^2 \le s_{56} &\le& (\sqrt{s}-\sqrt{s_{347}})^2,\\
\label{e3f}
m_3  \le E_3 &\le& \left[m_3^2+\lambda(s,m_3^2,m_4^2
(1+E_{\rm cut}/E_4^{\rm max}))/(4s)\right]^{1/2},\\
\label{e7f}
E'_{\rm cut} \le E_7 &\le& 
\frac{(\sqrt{s}-E_3)^2-E_3^2+m_3^2-m_4^2}
                           {2(\sqrt{s}-E_3-|{\bf p}_3|)},
\eea
where $E'_{\rm cut}$ is the photon energy cut transformed to the c.m.s.
of $u\bar{d}\gamma$ and $E_4^{\rm max}=\left[m_4^2+\right.$
$\left.\lambda(s,m_3^2,m_4^2)/(4s)\right]$ is the maximum of the $\bar{d}$
quark energy. We have neglected $E_{\rm cut}$ on the left hand side
of (\ref{s347}) and $E'_{\rm cut}$ in $E_4^{\rm max}$, which simplifies
the corresponding analytic expressions. The correct phase space boundaries 
are then restored by checking the condition $ E_7 \ge E_{\rm cut}$ in 
the c.m.s. numerically.
The spherical angles of (\ref{ps10f}) vary again in the full range.

The phase space parametrization convenient for the description of the
photon radiation off the $u$ quark or $\mu^-$ is obtained from 
(\ref{ps10f}) by a permutation of the final state momenta.

In order to improve the convergence of the phase space integration we
perform the following mappings. The Breit-Wigner shape of the $W^{\pm}$ 
resonances is taken into account by the mapping
\bea
\label{mapW}
s_{W}=\Gamma_W m_W \tan\left(\frac{\Gamma_W m_W}{N_{W}} x
+\psi_{\rm min}\right) + m_W^2,
\eea
where $N_W$ is the normalization factor, $N_{W}=\Gamma_W m_W /
\left(\psi_{\rm max} - \psi_{\rm min}\right)$, with $\psi_{\rm
min}=\arctan(s_{W}^{\rm min}-m_W^2)/(\Gamma_W m_W)$ and $\psi_{\rm
max}=\arctan(s_{W}^{\rm max}-m_W^2)/(\Gamma_W m_W)$.  The $1/t$ pole
due to the the neutrino exchange diagram (1) of Fig.~\ref{fig:1} is
mapped by transforming the polar angle of the virtual $W^+$ boson with
respect to the positron beam $\theta_W$ according to
\bea
\label{mapt}
 \cos{\theta_W}=(1-(1+\beta_W)r_W^{-x})/\beta_W,
\eea
where $\beta_W$ stands for for the velocity of the $W^+$ boson and 
$r_W=(1+\beta_W)/(1-\beta_W)$.

The $\sim 1/E_7$ peaking of the bremsstrahlung photon spectrum 
is eliminated by the mapping
\bea
\label{mape7}
E_7=E_7^{\rm min}\left(E_7^{\rm max}/E_7^{\rm min}\right)^{x},
\eea
where $E_7^{\rm min}$ and $E_7^{\rm max}$ are the lower and upper limit of the
photon energy.
The strong collinear peaking behavior of the squared matrix element 
of reaction (\ref{brems}) corresponding to the radiation off
the initial state positron $\sim 1/(1 - \beta\cos\theta_7)$ is eliminated 
by the mapping
\bea
\cos\theta_7=\frac{1}{\beta_e}\left(1-(1+\beta_e)/r_e^x\right)
\eea
with $r_e = (1+\beta_e)/(1-\beta_e)$ and $\beta_e=\sqrt{1-4m_e^2/s}$ 
being the velocity of the electron in the c.m. system. Similarly,
the collinear peaking related to the radiation off the initial state electron
$\sim 1/(1 + \beta\cos\theta_7)$ is dealt with the mapping
\bea
\cos\theta_7=\frac{1}{\beta_e}\left((1-\beta_e)r_e^x-1\right).
\eea
Finally the collinear and soft photon peaking corresponding to
radiation off the final state fermion must be mapped away, e.g. for
the $\bar{d}$ quark, $1/(p_4 \cdot p_7) \sim 1/(C_3-E_3)$ is
eliminated by the mapping
\bea
\label{mape3}
E_3=C_3-(C_3-m_3)\left(\frac{C_3-E_3^{\rm max}}{C_3-m_3}\right)^x,
\eea
where $C_3=\sqrt{s_{347}}/2+(m_3^2-m_4^2)/(2\sqrt{s_{347}})$.
In (\ref{mapW}--\ref{mape3}), $x$ denotes a random variable uniformly
distributed in the interval $[0, 1]$.

The phase space parametrizations (\ref{ps10i}) and (\ref{ps10f}) together
with the mappings (\ref{mapW}--\ref{mape3}) are implemented in a single
multichannel Monte Carlo program in a way described in \cite{EXCALIBUR}. 
We use five different channels corresponding to the photon radiation off 
each charged particle with equal weights.

In the soft photon limit, we can perform the integration over the photon phase
space analytically
\bea
\label{softint}
& &\left|{\rm d}\sigma_{\gamma}\right|_{|{\bf p}_7| < \omega}=
-\frac{1}{(2\pi)^3}\int_{|{\bf p}_7| < \omega}
\frac{{\rm d}^3p_7}{2E_7}
\left(  g_{\gamma l} {{p_1} \over {p_1 \cdot p_7}}
       - g_{\gamma l} {{p_2} \over {p_2 \cdot p_7}}\right.\nonumber\\
&+&   \left. g_{\gamma u} {{p_3} \over {p_3 \cdot p_7}}
       - g_{\gamma d} {{p_4} \over {p_4 \cdot p_7}}
       + g_{\gamma l} {{p_5} \over {p_5 \cdot p_7}}\right)^2
         {\rm d}\sigma_0= 
-\sum_{i,j=1}^5 g_{\gamma i}g_{\gamma j} I_{ij}^{\omega}.
\eea
The bremsstrahlung integrals $I_{ij}^{\omega}$, which are defined by
\bea
\label{iomega}
I_{ij}^{\omega}=\frac{1}{(2\pi)^3}\int_{|{\bf p}_7| < \omega}
\frac{{\rm d}^3p_7}{2E_7}\frac{p_i\cdot p_j}{(p_i\cdot p_7)(p_j\cdot p_7)}\;\;,
\eea
for $i \neq j$ may be found in Sect.~7 of \cite{tHV}. For 
$i=j$ we have
\bea
\label{iomegaii}
I_{ii}^{\omega}=\ln\frac{2\omega}{m_{\gamma}}-\frac{1}{\beta_i}
\ln\frac{1+\beta_i}{1-\beta_i},
\eea
where $\beta_i$ is the velocity of the radiating particle in the c.m.s.
and $m_{\gamma}$ denotes a fictitious mass of the photon.

\section{Numerical results}
\label{sec:5}

We now present our results for the Born cross section $\eeudmn$ and
the corresponding hard bremsstrahlung process.  We use the following
physical parameters: the gauge boson masses and widths $m_W=80.23$
GeV, $\Gamma_W=2.085$ GeV, $m_Z=91.1888$ GeV, $\Gamma_Z=2.4974$ GeV,
the fermion masses: $m_e=0.51099906$ MeV, $m_{\mu}=105.658389$ MeV,
$m_{\tau}=1777.05$ MeV, $m_u=5$ MeV, $m_d$=10 MeV, $m_s$=170 MeV,
$m_c$=1.3 GeV.  The electroweak standard
model couplings are parametrized by $\alpha_W=1/128.07$ and by the
electroweak mixing parameter $\sin^2\theta_W=0.22591$. The couplings
of the bremsstrahlung photon are parametrized by
$\alpha=1/137.0359895$ which means in practice that we multiply the
matrix element squared by the ratio $\alpha/\alpha_W$.

Our results have been thoroughly tested and checked against other
calculations.  The matrix elements have been checked against {\tt
MADGRAPH}~\cite{MADGRAPH}.  Moreover, as we already mentioned in
Sect.~\ref{sec:3}, we have checked the electromagnetic gauge
invariance of the matrix element of the bremsstrahlung process both
analytically and numerically.  The phase space integrals have been
checked against their asymptotic limits which have been obtained
analytically.

The total cross sections in the Born approximation  
are compared against {\tt EXCALIBUR}~\cite{EXCALIBUR}
in Tab.~\ref{tab:1}, with no cuts applied.\\[-13mm]
\begin{center}
\begin{table}[htb]
\caption{Born cross sections in pb (no cuts)}\smallskip
\label{tab:1} 
\begin{center}
\begin{tabular}{|r|l|l|}
\hline
 $E_{\rm cm}$ GeV &~~~~~$\sigma_{0}^{{\rm all}}$&
 ~~~$\sigma_{0}^{{\rm all}}$~\cite{EXCALIBUR}\\
\hline \hline
       162.5
&0.2685(0.4)&0.2688(3) \\
       180.0
&0.6612(0.9)&0.6616(7)  \\
       189.0
&0.7037(1.0)&0.7044(8)  \\
       500.0
&0.2810(0.5)&0.2817(5)  \\
      1000.0
&0.1078(0.3)&0.1079(2)  \\
      2000.0
&0.03736(1.2)&0.03748(8) \\
     10000.0
&0.002563(3)&0.002578(16) \\
\hline
\end{tabular}
\end{center}
\end{table}
\end{center}
In Tab.~\ref{tab:2}, we compare the Born cross sections $\sigma_0$ and the
corresponding hard bremsstrahlung cross sections $\sigma_{\gamma}$ with 
the results of \cite{DDRW} in the so called constant width scheme
and with phase-space integration restricted by the ``canonical'' cuts.
Let $l$, $q$, $\ga$, and ``beam'' denote
charged leptons, quarks, photons, and the beam (electrons or positrons), 
respectively, and $\theta(i,j)$ the angles between the particles $i$ and
$j$ in the c.m. system. Furthermore, $m(q,q')$ denotes
the invariant mass of a quark pair $qq'$. The ``canonical'' cuts then read: 
\beq
\begin{array}[b]{rlrlrl}
\theta (l,\mathrm{beam})> & 10^\circ, & \qquad
\theta( l, l^\prime)> & 5^\circ, & \qquad 
\theta( l, q)> & 5^\circ, \\
\theta (\ga,\mathrm{beam})> & 1^\circ, &
\theta( \ga, l)> & 5^\circ, & 
\theta( \ga, q)> & 5^\circ, \\
E_\ga> & 0.1\;\gv, & E_l> & 1\;\gv, & E_q> & 3\;\gv, \\
m(q,q')> & 5\;\gv\,.
\end{array}
\label{eq:canonicalcuts}
\eeq
Except for the additional angular cut between the charged leptons,
which is irrelevant for the reactions considered here anyway,
these cuts which exclude all collinear and infrared singularities coincide 
with those defined in \cite{Bardin}.\\[-10mm]
\begin{center}
\begin{table}[htb]
\caption{Cross sections\protect\footnotemark\ in fb with the ``canonical'' cuts
(\ref{eq:canonicalcuts})}\smallskip
\label{tab:2} 
\begin{center}
\begin{tabular}{|r|l|l||l|l|}
\hline
 $E_{\rm cm}$ GeV &~~~$\sigma_{0}$&
 ~~~$\sigma_{0}$~\cite{DDRW}&
 ~~~$\sigma_{\gamma}$&$\sigma_{\gamma}$ \cite{DDRW}\\
\hline \hline
       189.0 &   703.1(1)  &   703.5(3) &  223.2(8)    & 224.0(4)   \\
       500.0 &   237.4(1)  &   237.4(1) &   83.3(6)    &  83.4(3)   \\
      2000.0 &    13.96(1) &   13.99(2) &    7.05(8)   &   6.98(5)  \\
     10000.0 &     0.625(1)&   0.624(3) &    0.459(9)  &   0.457(6) \\
\hline
\end{tabular}
\end{center}
\end{table}
\end{center}
\footnotetext{Here we adopt the parameters of
\cite{DDRW}: $m_W=80.26$ GeV, $\Gamma_W=2.05$ GeV, $m_Z=91.1884$
GeV, $\Gamma_Z=2.46$ GeV, $\alpha=\alpha_W=1/128.89$ and
$\sin^2\theta_W = 1 - m_W^2/m_Z^2$.  The fermion masses play no role
in the presence of the cuts.}
Another test, which is very sensitive to the treatment of the infrared
and collinear singularities, is obtained by splitting the photon
radiation cross section (\ref{brems}) into a soft and a hard part
$\sigma_\gamma=\sigma_{{\rm s}} +\sigma_{{\rm h}}$ and checking the
independence of the separation cut. The soft part includes the photons
with $E_\gamma< \omega$ and is given by (\ref{softint}).  The hard
part then includes all photons with energies $E_\gamma >
\omega$. Furthermore, the total ``inclusive'' cross section is
$\sigma=\sigma_0+\sigma_{{\rm s}} +\sigma_{{\rm h}}$.  Since we have
not yet included the infrared (IR) singular virtual one-loop
corrections, $\sigma_{{\rm s}}$ only exists when it is IR regularized
in some way. Here we have chosen a small photon mass $m_\gamma =
10^{-6}\, \gv$.  We demonstrate the cut ($\omega$) independence of the
soft-hard splitting in Tab.~\ref{tab:3}. It is a measure of the
numerical stability of our calculation as well as a test for the
validity of the factorization into a radiation factor times the
non-radiative cross section (\ref{softint}).\\[-10mm]
\begin{center}
\begin{table}[htb]
\caption{Cut-independence of $\sigma_\gamma=\sigma_s+\sigma_h$. The photon
mass is $m_{\gamma}=10^{-6}GeV$.}\smallskip
\label{tab:3} 
\begin{center}
\begin{tabular}{|c|c|c|c|c|}
\hline
 $E_{\rm cm}$(GeV)& $\omega$ (GeV) & $\sigma_s$ (fb) & $\sigma_h$ (fb) & 
$\sigma_s +\sigma_h$ (fb) \\
\hline \hline
 189   & 0.001     & 202.6(2) & 1083(1)  & 1285  \\
       & 0.1       & 712.2(5) & 572.8(6) & 1285  \\
       & 1.0       & 967.0(7) & 319.3(3) & 1286  \\[1.5mm]
 500   & 0.001     &  42.53(4)& 528.3(1.0)& 570.8 \\
       & 0.1       & 247.3(2) & 322.7(6)  & 570.0 \\
       & 1.0       & 349.7(3) & 220.4(4)  & 570.0 \\[1.5mm]
2000   & 0.1       &  26.73(4)&  55.8(3)  &  82.5 \\
       & 1.0       &  40.75(8)&  42.4(3)  &  83.1 \\[1.5mm]
10000  & 0.1       &   1.302(5) & 4.85(6) &   6.15 \\
       & 1.0       &   2.265(8) & 3.84(6) &   6.10 \\
\hline
\end{tabular}
\end{center}
\end{table}
\end{center}
\begin{center}
\begin{table}[htb]
\caption{Mass dependence of cross sections (in fb; without cuts, except for
$E_{\gamma}>0.1$ GeV) for different final states}
\label{tab:4} 
\begin{center}
\begin{tabular}{|r|l|l|l||l|l|l|}
\hline
 $E_{\rm cm}$ GeV &$\sigma_{0}(u \bar{d}
\mu^- \bar{\nu}_{\mu})$&
 $\sigma_{0}(c \bar{s}
\mu^- \bar{\nu}_{\mu})$&
 $\sigma_{0}(u \bar{d}
\tau^- \bar{\nu}_{\tau})$&
 $\sigma_{\gamma}(u \bar{d}
\mu^- \bar{\nu}_{\mu})$&$\sigma_{\gamma}(c \bar{s}
\mu^- \bar{\nu}_{\mu})$&
 $\sigma_{\gamma}(u \bar{d}
\tau^- \bar{\nu}_{\tau})$\\
\hline \hline
       189.0   
& 704.1(4)  & 703.8(4) & 703.5(4) 
& 573.4(4)  & 525.2(4) & 522.6(4) \\
       360.0   
& 422.0(2)  & 421.8(2) & 421.5(2)
& 448.5(4)  & 418.4(4) & 414.1(4)  \\
       500.0   
& 281.0(2)  & 280.9(2) & 281.0(2)
& 322.8(4)  & 302.0(4) & 298.1(3)  \\
      2000.0   
& 37.33(4)  & 37.32(4) & 37.32(4)
& 56.48(27) & 53.19(25)& 52.67(13) \\
\hline
\end{tabular}
\end{center}
\end{table}
\end{center}
\newpage
Finally, in Tab.~\ref{tab:4} we illustrate the mass dependence of the
cross sections of related channels. The replacements $ud \ra cs$ and
$\mu \ra \tau$ lead to comparable effects. While the Born cross
sections remain practically unchanged the hard bremsstrahlung cross
section, for the energy cut $E_{\gamma}>0.1$ GeV, changes by about
$-9$\% (189 GeV) to about $-6$\% (2 TeV).
\begin{figure}[hb]
\resizebox{0.8\textwidth}{!}{%
  \includegraphics{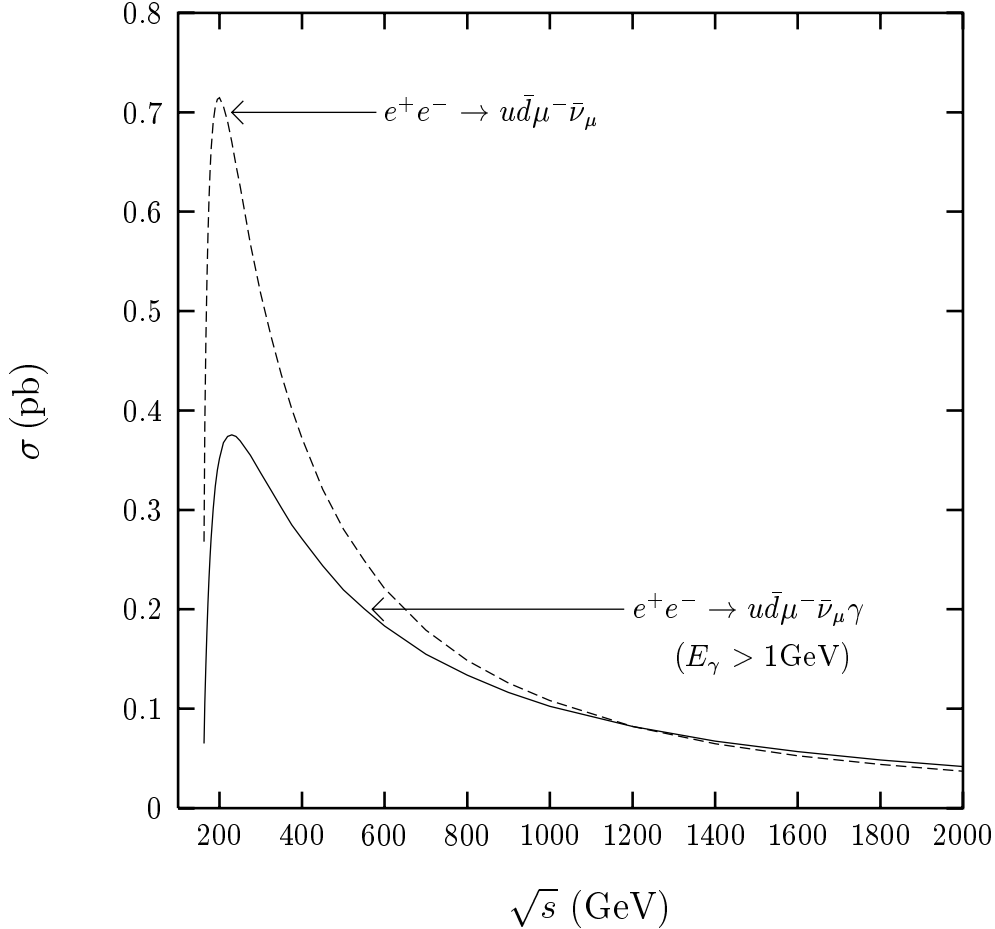}
}
\caption{The energy dependence of the total cross sections
                  of reactions (\ref{Born}) and (\ref{brems}).}
\label{fig:2}
\end{figure}

The energy dependence of the total cross sections $\sigma_0$ and
$\sigma_\gamma$ of reactions $e^+ e^- \ra u \bar{d} \mu^-
\bar{\nu}_{\mu}$ and $e^+ e^- \ra u \bar{d} \mu^-
\bar{\nu}_{\mu}\:\gamma$, respectively, is shown in Fig.~\ref{fig:2}. 
The hard bremsstrahlung cross section has been calculated with the
photon energy cut of $E_{\gamma} = 1$ GeV.

At this point we would like to address the problem of $SU(2)$
gauge-symmetry violation caused by introducing the constant widths
$\Gamma_W$ and $\Gamma_Z$ in a more quantitative way. For this
purpose, in Fig.~\ref{fig:3a}, we compare the $\eeudmn$ cross sections
obtained when i) including all diagrams, ii) including only
$WW$-diagrams and iii) assuming the creation of an on-shell
$W^{\pm}$-pair and the subsequent decays $W^-
\ra \mu^- \bar{\nu}_{\mu}$ and $W^+ \ra u \bar{d}$ ($e^+ e^- \ra W^+W^- \ra 
u \bar{d} \mu^- \bar{\nu}_{\mu}$). Cases i) and ii) for the
corresponding bremsstrahlung reaction with the photon energy cut
$E_{\gamma} > 1$ GeV are plotted in Fig.~\ref{fig:3b}.  Although the
constant width prescription violates unitarity by spoiling the gauge
cancellations in both cases i) and ii), the unitarity violation is
much stronger in case ii) where we have neglected the non
double-resonant diagrams. In case i) the effect is practically
negligible, at least in the energy range presented in
Fig.~\ref{fig:3a}. This observation relies on the comparison with the
results of \cite{DDRW}. Our results which were calculated in the
linear 

\newpage

\rput(7,-4.7){\scalebox{0.8 0.8}{\epsfbox{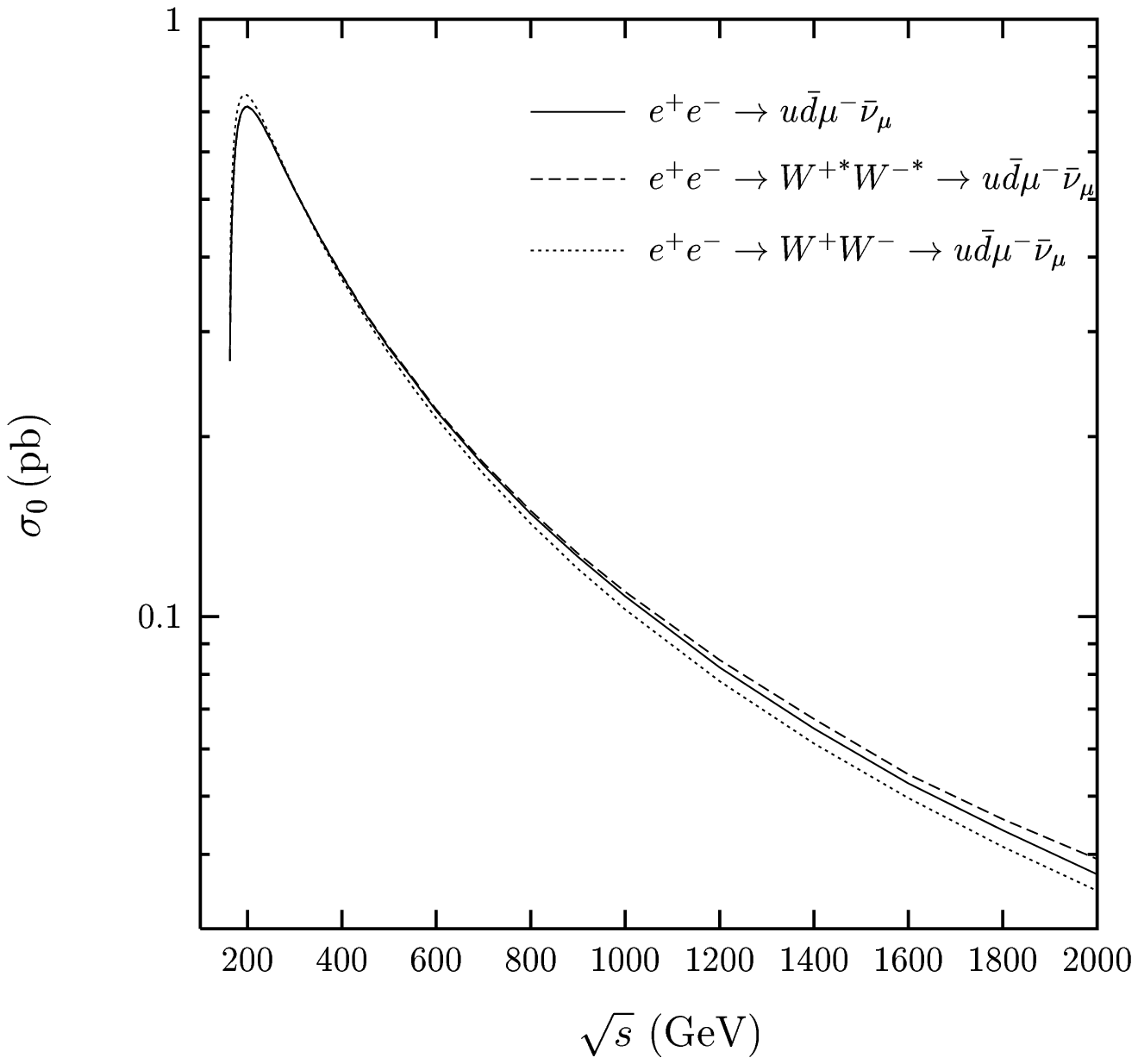}}}

\vspace*{9.2cm}

\begin{figure}[!ht]
\caption{Lowest order cross sections obtained by
including i) all diagrams, ii) only diagrams with $W^* W^*$ intermediate 
states and iii) production and decay of on-shell $W$-pairs (zero-width 
approximation).}
\label{fig:3a}
\end{figure}

\rput(7,-4.7){\scalebox{0.8 0.8}{\epsfbox{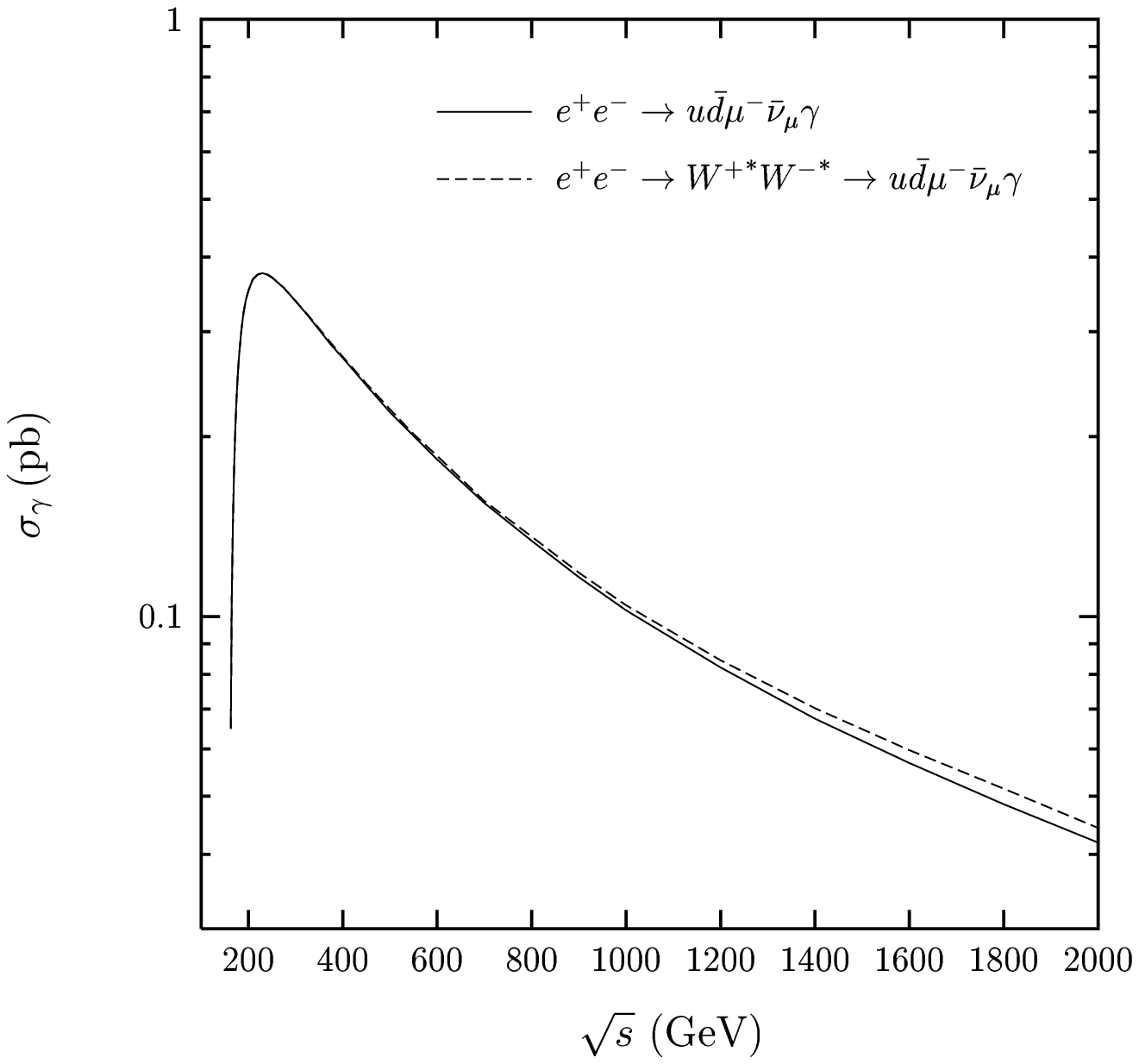}}}

\vspace*{9.2cm}

\begin{figure}[!h]
\caption{Bremsstrahlung cross sections for $E_{\gamma} > 1$ GeV
obtained by including i) all diagrams, ii) only diagrams with $W^* W^*$ 
intermediate states.}
\label{fig:3b}
\end{figure}

\newpage

gauge agree within statistical errors with those of
\cite{DDRW} which were obtained in a nonlinear gauge in the so
called complex-mass scheme that preserves the Ward identities. In
Figs.~\ref{fig:3a} and \ref{fig:3b} we observe that the
double-resonant approximation provides a good approximation of the
complete calculation for c.m. energies from threshold up to about 250
GeV. However, it already deviates by 0.5\% (more than 1\% for reaction
(\ref{brems})) at $\sqrt{s}=500$ GeV and by about 2\% at $\sqrt{s}=1$
TeV.  On the other hand, we see that the zero width approximation,
which is gauge invariant by definition, deviates from the complete
tree level calculation by almost 18\% at $\sqrt{s}=165$ GeV, by 4.3\%
at $\sqrt{s}=200$ GeV and by $-4.5$\% at $\sqrt{s}=1$ TeV.

\begin{figure}[htb]
\resizebox{0.8\textwidth}{!}{%
  \includegraphics{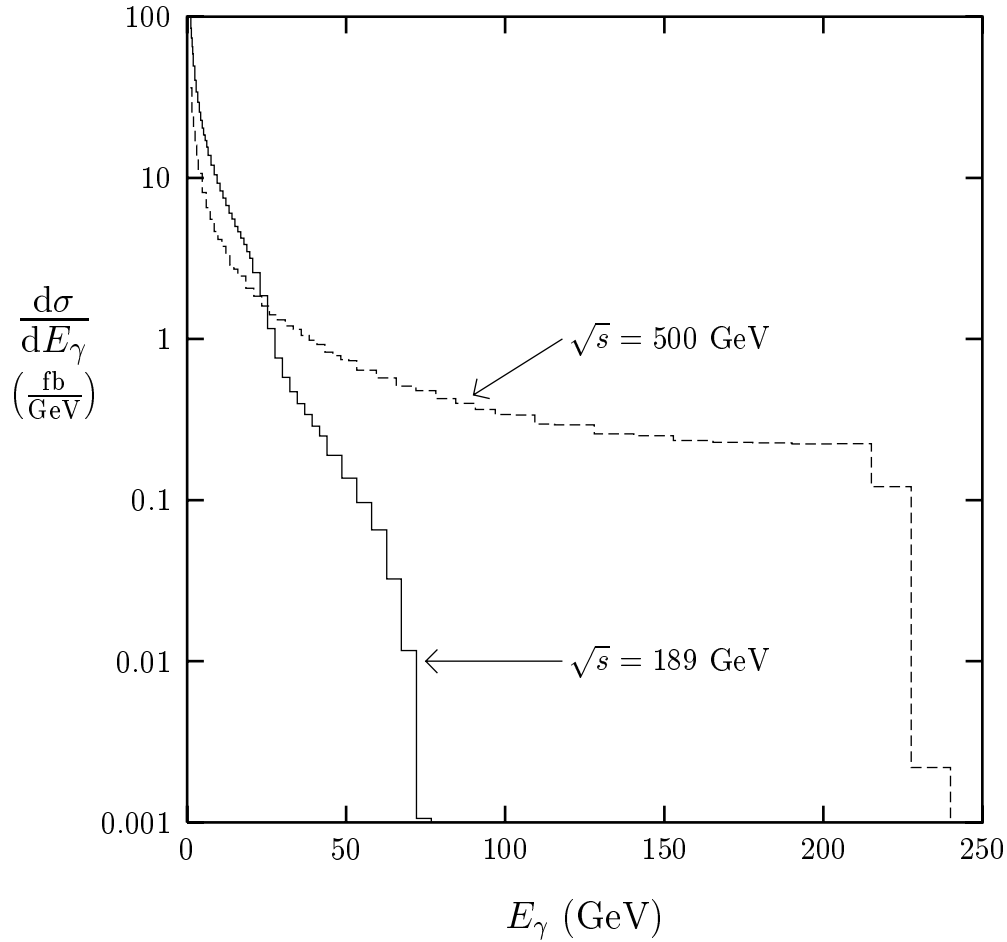}
}
\caption{The photon spectra at $\sqrt{s}=189$ GeV and
$\sqrt{s}=500$ GeV.}
\label{fig:4}
\end{figure}

The photon spectra at $\sqrt{s}=189$ GeV and $\sqrt{s}=500$ GeV are
shown in Fig.~\ref{fig:4}. We see that the spectra are relatively
soft, with a substantial fraction of events having photon energies of
$O(\Gamma_W)$. A bump of the 189 GeV spectrum at $E_{\gamma} \sim 25$
GeV reflects the $W$-pair production threshold. We finally plot ${\rm
d}\sigma /{\rm d}m^2_{34}$ at $\sqrt{s}=189$ GeV as a function of the
invariant mass of the $u\bar{d}$ pair $m_{34}$ in Fig.~\ref{fig:5}.

\begin{figure}[htb]
\resizebox{0.8\textwidth}{!}{%
  \includegraphics{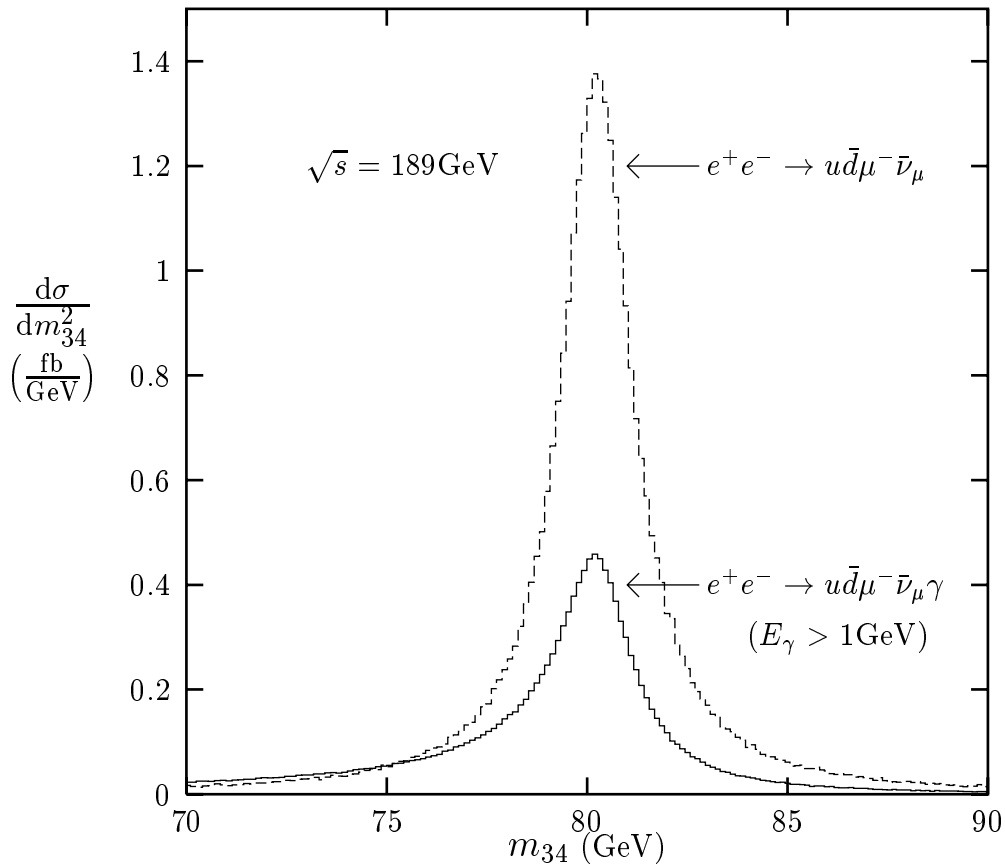}
}
\caption{The differential cross section ${\rm d}\sigma /{\rm
d}m^2_{34}$ at $\sqrt{s}=189$ GeV versus the invariant mass of the
$u\bar{d}$ pair $m_{34}$. 
The upper curve corresponds to the tree
level reaction (\ref{Born}) and the lower curve to the bremsstrahlung
reaction (\ref{brems}) with $E_{\gamma} > 1$ GeV.}
\label{fig:5}
\end{figure}

\bigskip

\section{Conclusions and outlook}
\label{sec:6}

We have presented an efficient method for calculating photon radiation
cross sections for massive fermions. For the collinear region such
finite mass calculations provide important tests for Monte Carlo
generators which work with massless fermions. We have studied a
complete signal plus background process with all possible real photon
emission diagrams for the interesting channel $\eeudmn$. This channel is
particularly suited for a detailed investigation of effects related to
final state photon emission, since the muons appear well separated
from photons in the detectors. In particular it seems to be
interesting to study the influence of final state radiation on the $W$
mass measurement via this channel. In addition at a high luminosity
linear collider, like TESLA, one could study the quark mass effects
due to the different quark flavor channels in $e^+ e^- \ra 
\mu^- \bar{\nu}_{\mu}\,+\,{\mathrm{hadrons}}$. Of particular interest
would be a detailed investigation of the single top production
channel $e^+e^- \ra t \bar{b} \mu^- \bar{\nu}_{\mu}$ which will be discussed
in a forthcoming paper.

\bigskip

{\bf Acknowledgment}

One of us (K.K.) would like to thank H. Czy\.z and J. S\l adkowski
for discussions. Furthermore, we thank T. Riemann for helpful
discussions and for carefully reading the manuscript.


\bigskip

\begin{thebibliography}{99}
\bibitem{Bardin} 
D.~Bardin et al., \textit{Event generators for $WW$ physics}, 
in \textit{Physics at LEP2}, CERN 96-01 (1996)
(G.~Altarelli, T.~Sj\"o\-strand, F.~Zwirner, eds.),
vol.~2, pp.~3-353

\bibitem{LEP}
D. Karlen, \textit{Experimental Status of the Standard Model},
in Proc. of \textit{29th International Conference on High-Energy
Physics} (ICHEP 98), 23-29
July 1998, Vancouver, Canada 

\bibitem{rc} M. B\"ohm et al., Nucl. Phys. B \textbf{304} (1988) 463;\\
             J. Fleischer, F. Jegerlehner, M. Zra\l ek, Z. Phys. C \textbf{42}
             (1989) 409
\bibitem{KZ} K.~Ko\l odziej, M.~Zra\l ek, Phys. Rev. D \textbf{43} (1991) 3619
\bibitem{hb} W. Beenakker, K. Ko\l odziej, T. Sack, Phys. Lett. B
             \textbf{258} (1991) 469;\\
             W. Beenakker, F.A. Berends, T. Sack, Nucl. Phys. B 
             \textbf{367} (1991) 287;\\
             J. Fleischer, K. Ko\l odziej, F. Jegerlehner,
             Phys. Rev. D \textbf{47} (1993) 830
\bibitem{SD} S. Dittmaier, CERN-TH/98-336, {\tt hep-ph/9811434}
\bibitem{AV} A. Vicini, Acta Phys. Pol. B \textbf{29} (1998) 2847
\bibitem{AW91} A. Aeppli and D. Wyler, Phys. Lett. B \textbf{262} (1991) 125;\\
               A. Aeppli, doctoral thesis, Universit\"at  Z\"urich (1992)
\bibitem{GJ} G.J. van Oldenborgh, P.J. Franzini, A. Borrelli, Comput.
             Phys. Commun. \textbf{83} (1994) 14
\bibitem{JF} J. Fujimoto, et al., Nucl. Phys. (Proc. Suppl.) 
             \textbf{37}B (1994) 169
\bibitem{CM} F. Caravaglios, M. Moretti, Z. Phys. C \textbf{74} (1997) 291
\bibitem{DDRW} A. Denner, S. Dittmaier, M. Roth, D. Wackeroth,
               BI-TP 99/10, PSI-PR-99-12, {\tt hep-ph/9904472}
\bibitem{BD} W. Beenakker, A. Denner, Acta Phys. Pol. B 
             \textbf{29} (1998) 2821
\bibitem{BBC} W. Beenakker, F.A. Berends, A.P. Chapovsky,
             Phys. Lett. B \textbf{435} (1998) 233
\bibitem{vegas} G.P. Lepage, J. Comp. Phys. \textbf{27} (1978) 192
\bibitem{EXCALIBUR} F.A. Berends, R. Pittau, R. Kleiss, Nucl. Phys. B 
              \textbf{424} (1994) 308; \\
              Comput. Phys. Commun. \textbf{85} (1995) 437 
\bibitem{tHV} G.~'t Hooft, M. Veltman, Nucl. Phys. B \textbf{153} (1979) 365
\bibitem{MADGRAPH} T. Stelzer, W.F. Long, Comput. Phys. Commun. 
               \textbf{81} (1994) 357;\\
                E. Murayama, I. Watanabe, K. Hagiwara, KEK report 91--11,
                   1992
\end{thebibliography}
\end{document}